\documentclass{aa}  	

\usepackage{graphicx}
\usepackage{amsmath}	
\usepackage{amssymb}	
\usepackage{soul}

\usepackage{txfonts}
\newcommand{\Msun}{$M_{\odot}$}

\newcommand{\kms}{$\mbox{km s}^{-1}$}
\newcommand{\lambdaRe}{$\lambda_\mathrm{Re}$}
\newcommand{\epsE}{$\epsilon_\mathrm{e}$}

\begin{document}

   \title{Nuclear angular momentum of early-type galaxies \\ hosting nuclear star clusters
   		\thanks{Based on observation collected at the ESO Paranal La Silla Observatory, Chile, Prog. ID 092.B-0892, PI  Lyubenova and ID 380.B-0530, PI Infante.}
   }

   \author{Mariya Lyubenova
   	         \inst{1}
          \and
          Athanassia Tsatsi
                   \inst{2}
          }

   \institute{ESO, Karl-Schwarzschild-Str. 2, D-85748 Garching bei  M\"unchen, Germany\\
		   \email{mlyubeno@eso.org}
                    \and
                    Max Planck Institute for Astronomy, K\"onigstuhl 17, D-69117 Heidelberg, Germany
             }

   \date{Received: \dots ; accepted: \dots}

  \abstract
  {Nucleation is a common phenomenon in all types of galaxies and at least 70\% of them host nuclear star clusters (NSCs) in their centres. Many of the NSCs co-habit with super-massive black holes and follow similar scaling relations with host galaxy properties.  NSCs, unlike black holes, preserve the signature of their evolutionary path imprinted onto their kinematics and stellar populations. Thus their study provides us with important information about the formation of galactic nuclei.}
   {In this paper we explored the angular momentum of the nuclei of six intermediate mass ($9.7 > \log(M_{dyn}/M_{\sun}) > 10.6$) early-type galaxies in the Fornax cluster that host NSCs. Our goal was to derive a link between the nuclear angular momentum and the proposed formation scenarios of NSCs.}
   {We used Adaptive Optics assisted IFU observations with VLT/SINFONI to derive the spatially resolved stellar kinematics of the galaxy nuclei. We measured their specific stellar angular momenta \lambdaRe\/, and compared these with Milky Way globular clusters and $N$-body simulations of NSC formation.}
   {We found that all studied nuclei exhibit varied stellar kinematics. Their \lambdaRe\/ and ellipticities are similar to Milky Way globular clusters (GCs). Five out of six galaxy nuclei are consistent with the \lambdaRe\/--\epsE\/ of simulated NSCs embedded in a contaminating nuclear bulge that have formed via the in-spiralling and merging of GCs. }
   {It has previously been suggested that the NSCs in higher mass galaxies, like the ones studied in this paper, form via dissipational sinking of gas onto the galactic nuclei with hints that some might also involve the merger of GCs. Here we showed that we cannot exclude the pure GC merging scenario as a viable path for the formation of NSCs.}


   \maketitle
%

\section{Introduction}
\label{sec:intro}

Observations made over the last few decades have shown that nucleation is a common phenomenon, with at least $70\%$ of galaxies over a broad mass range hosting a nuclear star cluster (NSC) in their photometric and kinematic centres \citep[e.g.][]{boeker02, cote06, neumayer11,turner12,denbrok14,georgiev14}. Often these NSCs co-exist with supermassive black holes \citep[SMBH, e.g.][]{graham09, georgiev16}. The relationship between the mass of these central massive objects (CMOs) and the properties of their host galaxies is likely to be fundamental as it connects quantities that differ by several orders of magnitude \citep[e.g.][]{ferrarese06}. It is still debated whether this relationship is physical through feedback from the SMBH on the host galaxy \citep[e.g.][]{silk98}, or statistical through many subsequent mergers of galaxies and their black holes \citep{jahnke11}. The answer might well come from the study of the formation and evolution of NSCs, which, unlike black holes, preserve their evolutionally history imprinted onto their stellar populations and kinematics.

\begin{table*}
\begin{center}
\caption{\label{tab:sample} Basic properties of the 6 galaxies studied in this paper. Column (1) lists the \citet{ferguson89} catalogue designation and (2) the galaxies common names. Column (3) give the galaxies morphological type as listed in \citet{ferguson89}. The distances (4) are taken from \citet{blakeslee09} and are based on surface brightness fluctuations. The dynamical masses (5) are estimated using $M_{dyn} = 5.0~R_{eff} \sigma_{eff}^2 /G$  \citep{cappellari06}.  $R_{eff}$ (6) is taken from \citet{ferguson89}. The central velocity dispersions (7) are taken from \citet{kun00}, except for FCC277 and FCC310, for which they are taken from \citet{wegner03}. In order to use these as mass estimators in (5) we corrected them using the aperture corrections presented in \citet{kin_dyn_0}. Column (8) lists the total $B$-band magnitude extracted from HyperLeda.
}
\begin{tabular}{c c c c c c c c }
\hline
\hline
FCC ID  & Name& Type & Distance & $ \log(M_{dyn}) $ &  $R_{eff}$ & $\sigma$ & $B_T$  \\
  & & & Mpc & $M_{\sun}$ & \arcsec & \kms & $mag$  \\
(1) & (2) & (3) & (4) & (5) & (6) & (7) & (8) \\
\hline

FCC47	& NGC1336	&	E4			& 18.3	& 10.4	&	30.0	&	96.0	&	13.3		\\
FCC148	& NGC1375	&	S0			& 19.9	& 10		&	26.9	&	56.0	&	13.5		\\	
FCC170	& NGC1381	&	S0			& 21.9	& 10.6	&	12.9	&    153.0	&	12.9		\\
FCC177	& NGC1380A	&	S0			& 20.9	& 9.7		&	12.6	&	55.0	&	13.6		\\
FCC277	& NGC1428	&	E5			& 20.7	& 9.9		&	10.1	&	81.7	&	14.1		\\
FCC310	& NGC1460	&	SB0			& 19.9	& 10		&	25.8	&	60.4	&	13.7		\\
\hline
\hline
\end{tabular}
\label{default}
\end{center}
\end{table*}%

Two main scenarios for NSC formation have been proposed. The first one involves the dry merging of globular clusters that migrated towards the centre of the galaxy due to dynamical friction \citep[e.g.][]{tremaine75}. In the second scenario NSCs form {\it in situ} via dissipational sinking of gas onto the galactic nucleus and subsequent star formation \citep[e.g.][]{mihos94}. Various studies showed that it is difficult to unambiguously reconcile the proposed theoretical models with the plethora of observational phenomena that NSCs exhibit. For example, \citet{turner12} studied the photometric structural parameters of a large sample of NSCs in the Fornax and Virgo clusters and suggested that nuclei in high-mass early-type galaxies most likely grow through gas accretion triggered by wet mergers. At lower masses they suggested that the dominant mechanism is probably the infall and merging of star clusters with a possible hybrid population where both mechanisms work simultaneously.  \citet{spengler17} investigated the structural parameters of nuclei in the Virgo cluster and found that the most massive host galaxies tend to have flatter nuclei, suggesting that they may be formed predominantly through dissipative processes that can induce flattening and rotation.

When looked up close spectroscopically and with adaptive optics galactic nuclei display even larger variety: many NSCs contain multiple stellar populations and some are embedded in stellar or gaseous disks \citep[e.g.][]{walcher05,rossa06,seth06,seth08,barth09,az13}. These  complex properties have been explained with the help of numerical simulations involving both the merging of star clusters and gas dissipation \citep[e.g.][]{hartmann11,guillard16}. However, \citet{tsatsi17a} showed that pure globular cluster merging simulations \citep{antonini12,perets14} provide a good explanation of the observed stellar kinematics substructures of the closest NSC, the one in the Milky way \citep{feldmeier14}. 

In \citet{az13} we showed that even though the NSC in an early-type galaxy can have $|V|/\sigma < 1$, detailed dynamical modelling revealed that a significant stellar angular momentum is still preserved in the nucleus. Our first orbit-based dynamical model of a NSC revealed that co- and counter-rotating orbits are simultaneously needed to reproduce the observed stellar kinematics of the nuclear region in FCC\,277. This counter rotation is indicative of some merger event with orbital angular momentum opposite to the host galaxy. 

The increasing amount of excellent quality adaptive optics assisted integral-field spectroscopic data allows us to look at galactic nuclei in a way that was earlier possible only for the large scale structure of galaxies. Even though photometrically regular, the 2D stellar kinematics maps of early-type galaxies (ETGs) often exhibit complex structures, with mis-aligned and counter-rotating components \citep[e.g.][]{krajnovic11}. Based on their specific stellar angular momentum and ellipticity ETGs separate in two sub-groups -- slow- and fast-rotators \citep{emsellem07,emsellem11}. Slow rotator ETGs are usually more massive and mildly triaxial, while fast rotators are less massive, fainter, and oblate axisymmetric in shape \citep{emsellem11,weijmans14}. Cosmological hydrodynamical simulations suggest that slow rotators may result from dry mergers within the red sequence and that fast rotators likely result from gas-rich mergers \citep{naab14}.

With the goal to shed further light on the formation paths of NSCs, in this paper we present our exploration of the nuclear angular momentum of early-type galaxies hosting NSCs using a similar formalism that is typically used for the host galaxies. We present new observations of 5 more nuclei of early-type galaxies in the Fornax cluster (see Table~\ref{tab:sample} for their basic properties). We derived their stellar kinematics and together with our earlier data for FCC\,277, measured the specific stellar angular momentum \lambdaRe\/ of the final sample of 6 NSCs. We compared these measurements with the ones of globular clusters in the Milky Way \citep{kamann18} as well as with simulated dry merger remnants from \citet{antonini12} and \citet{perets14}.


\section{Observations and data reduction}
\label{sec:obs_datared}

\subsection{Sample selection}
\label{sec:sample}

In this paper we analysed the nuclear stellar kinematics of six early-type galaxies in the Fornax cluster.  Our pilot study of the nucleus of the galaxy FCC\,277, presented in \citet{az13}, made use of Natural Guide Star assisted adaptive optics (AO) observations therefore the main selection criteria for that study was the availability of a suitable bright star nearby the nucleus. We compiled our sample of 5 new targets based on two main selection criteria -- the brightness of the NSCs and their contrast with respect to the main body of the galaxy, as derived by \citet{turner12}. These criteria ensured the collection of high-quality Laser Guide Star AO assisted IFU data within reasonable amounts of observing time. In Table~\ref{tab:sample} we listed the basic properties of the selected host galaxies and in Table~\ref{tab:lambda_all} some basic properties of the NSCs can be found.

\subsection{Observations}
\label{sec:obs}

The observations were carried out in Service Mode between 21 October and 29 November 2013 with the VLT/SINFONI (programme ID 092.B-0892, PI M. Lyubenova). In order to determine the spatially resolved properties of the NSCs we used the Adaptive Optics system in Laser Guide Star (LGS) mode. Due to the brightness and the point like nature of our NSCs, we used them as tip-tilt stars for the LGS system. The observing conditions were always very good, with clear sky and guide probe seeing $<$0.8\arcsec. We obtained all observations using SINFONI's $K$-band grating (1.95 -- 2.45 $\mu$m) that gives a spectral resolution R$\sim$3500 (6.2 \AA\/ FWHM as measured on sky lines).

Our data cover the central 3\arcsec $\times$ 3\arcsec of each galaxy, with a spatial sampling of 0\farcs05 $\times$ 0\farcs10. We observed the galaxies in observing blocks (OBs) consisting of OOSOOSOOS (O=Object, S=Sky) sequence where we spent 6$\times$300 s on source and 3$\times$300 s on sky. This is an optimal configuration for minimising near-IR sky line residuals in SINFONI observations of extended sources. Each object exposure was dithered with respect to the previous ones with 0\farcs15 to reject bad pixels. The sky pointings were offset to North and East of each galaxy nucleus with amplitudes of 110\arcsec (FCC\,47), 71\arcsec (FCC\,148), 60\arcsec (FCC\,170), 56\farcs6 (FCC\,177), and 71\arcsec (FCC\,310).

FCC\,170 is the brightest galaxy in our sample, therefore to reach the required depth of observations we needed two such OBs with a total on source time of 1 hour. For FCC\,47, FCC\,148, FCC\,177, and FCC\,310 we used 5 OBs, configured in a similar manner. For these observations the total on source time was 2.5 hours. In order to remove the absorption features originating in the Earth’s atmosphere, after each observing block and at a similar airmass, we observed a B dwarf star to act as a telluric reference star. This kind of hot stars are particularly suited due to their relative lack of intrinsic absorption lines, apart of the Brackett\,$\gamma$ absorption line at 2.166\,$\mu$m, and their continuum in the $K$-band is well approximated by the Rayleigh-Jeans part of the black body spectrum, associated with their effective temperature.

\subsection{Data reduction}
\label{sec:data_red}

The data reduction of FCC\,47, FCC\,148, FCC\,170, FCC\,177, and FCC\,310 was similar to the data reduction of FCC\,277 presented in \citet{az13}, however with a few  exceptions. We used a newer ESO SINFONI pipeline version 2.5.2 to perform the basic data reduction and obtain sky-subtracted data cubes, as well as their corresponding telluric stars. Unlike FCC\,277, here we did not use the pipeline to combine the individual exposures within one OB because we noticed that the centres of the galaxy nuclei did not always fall on the same $(x,y)$ coordinates along the wavelength direction of each data cube. 

To analyse this issue we reconstructed images, consisting of the median flux in every 10 -- 20 wavelength pixels (corresponding to 24.5 -- 49 \AA\/ width) of each sky-subtracted data cube of one on-source exposure. We estimated the median luminosity weighted centre of each image and found that these are offset by typically 1--2 spatial pixels (0\farcs05 -- 0\farcs10) in $x$ and $<$1 pixel (0\farcs05) in $y$ direction, the strongest offset being usually red-wards of 2.2~$\mu$m. We note that these shifts are rather small and we could not identify their source during the previous data reduction steps. We speculate that these might be due to residuals from the applied differential atmospheric correction. We were able to notice them in our data due to their very good quality. Our observations were taken in excellent sky conditions and the use of the LGS fed AO allowed for a considerable Strehl ratio. As a result, our NSCs are resolved and we cannot simply neglect these shift. Therefore we corrected for them by shifting each monochromatic cube plane with respect to the centre of the first plane in each individual exposure sky-subtracted data cube, using a bilinear interpolation method. After that we computed and applied an optimised telluric correction for each of these cubes. Finally, we combined the data cubes from all available exposures of a given object (30 cubes for each object, except FCC\,170 which had 12) using a biweight mean and obtained estimates of the error spectra as the standard deviations of these.

Our data reach a minimum S/N$\sim$~40  {\it per spaxel} at the effective radius of the NSCs in FCC\,148, FCC\,170, and FCC\,177. In FCC\,47 and FCC\,310 the minimum S/N at these radii is $\sim$~20. To derive reliable stellar kinematics with the method that we used (See Sect.~\ref{sec:stellar_kin}) a minimum S/N$\sim$40 is needed. Therefore, we spatially binned our data to achieve this threshold also further out of the NSCs effective radii with the help of the Voronoi 2D binning method of \citet{cc03}.

\subsection{Spatial resolution}
\label{sec:spat_res}

%
\begin{figure}
\includegraphics[width=\linewidth, trim=0cm 0.3cm 1.2cm 0.5cm clip]{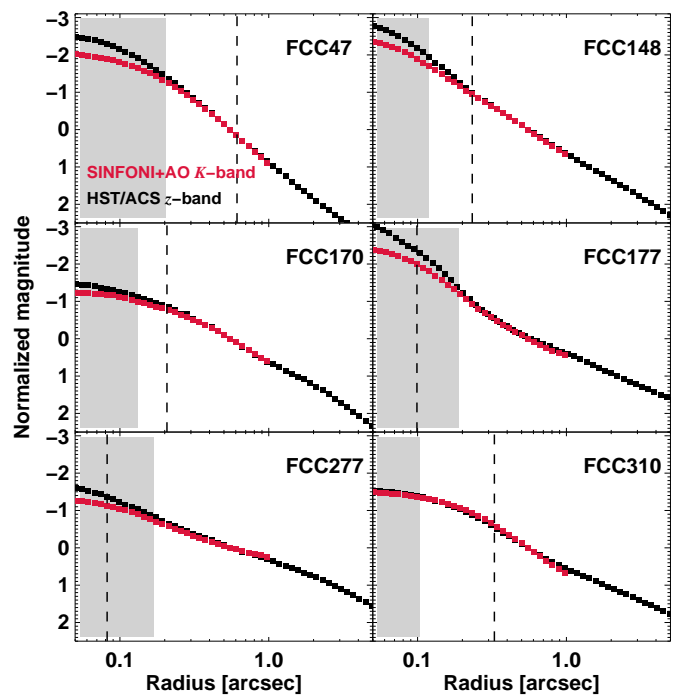}
\caption{\label{fig:nsc_lp} Comparison between the light profiles of the galaxies in our sample from our ground-based plus AO imaging and space based one. The red symbols denote our SINFONI+AO $K$-band reconstructed images, the black symbols  -- the HST/ACS $z$-band imaging. The light profiles were normalised in the region $ 0.5 \arcsec < Radius < 0.7 \arcsec$. The vertical dashed lines in each panel indicate the effective radius in $z$-band of each NSC, as derived by \citet{turner12}. The shaded area indicates the PSF $FWHM$ of the SINFONI data.}
\end{figure}

The goal of our study is to measure the 2-d stellar kinematics of the nuclear regions of galaxies hosting nuclear star clusters. Thus it is important to know how well we resolve their spatially extended properties. In Figure~\ref{fig:nsc_lp} we show a comparison between the derived isophotal light profiles from the HST and SINFONI. The black points refer to the HST/ACS  $z$-band imaging (Ferrarese et al. {\it in preparation}). The red points denote our SINFONI+AO $K$-band reconstructed images, and the vertical dashed lines indicate the effective radii of the hosted NSCs. The SINFONI profiles are in general less steep in the nuclear regions and we attribute this to the lower spatial resolution of the SINFONI data as compared to the ACS. We note, however, that the sizes of NSCs increase with wavelength \citep[e.g.][]{georgiev14,carson15}. This effect may well contribute to the increased width of the SINFONI profiles.

During the observing runs point-spread function (PSF) calibrator stars were not observed thus we used a different method to assess the spatial resolution of our Adaptive Optics fed SINFONI data, which also takes into account the effect of the re-centring of the monochromatic planes of the data cubes, described in Sect.~\ref{sec:data_red}.\footnote{We note that telluric stars are not useful for this purpose as these calibration observations, as a rule, are taken without the AO system.} We achieved this by convolving the ACS $z$-band images of each galaxy \citep[using a Tiny Tim PSF,][$\sim0.1\arcsec$]{krist95} with a given Gaussian PSF until it matched the light distribution in the reconstructed SINFONI images. The grey shaded areas in Figure~\ref{fig:nsc_lp} illustrate the $FWHM$ of these so called PSFs. We note that these are conservative upper limits due to the increase of NSC sizes with wavelength mentioned above.

In majority of the galaxies (four out of six) the $FWHM$ of our final spatial resolution is smaller than the effective radii of the hosted NSCs. In three cases, FCC\,148, FCC\,170, and FCC\,310, the spatial resolution of SINFONI is very similar to the HST. Thus we conclude that we are able to spatially resolve the stellar kinematics of the NSCs in FCC\,47, FCC\,148, FCC\,170, and FCC\,310.
%

\section{Stellar kinematics}
\label{sec:stellar_kin}
%
\begin{figure*}
\includegraphics[width=\linewidth, trim=0.3cm 6cm 0.4cm 0, clip]{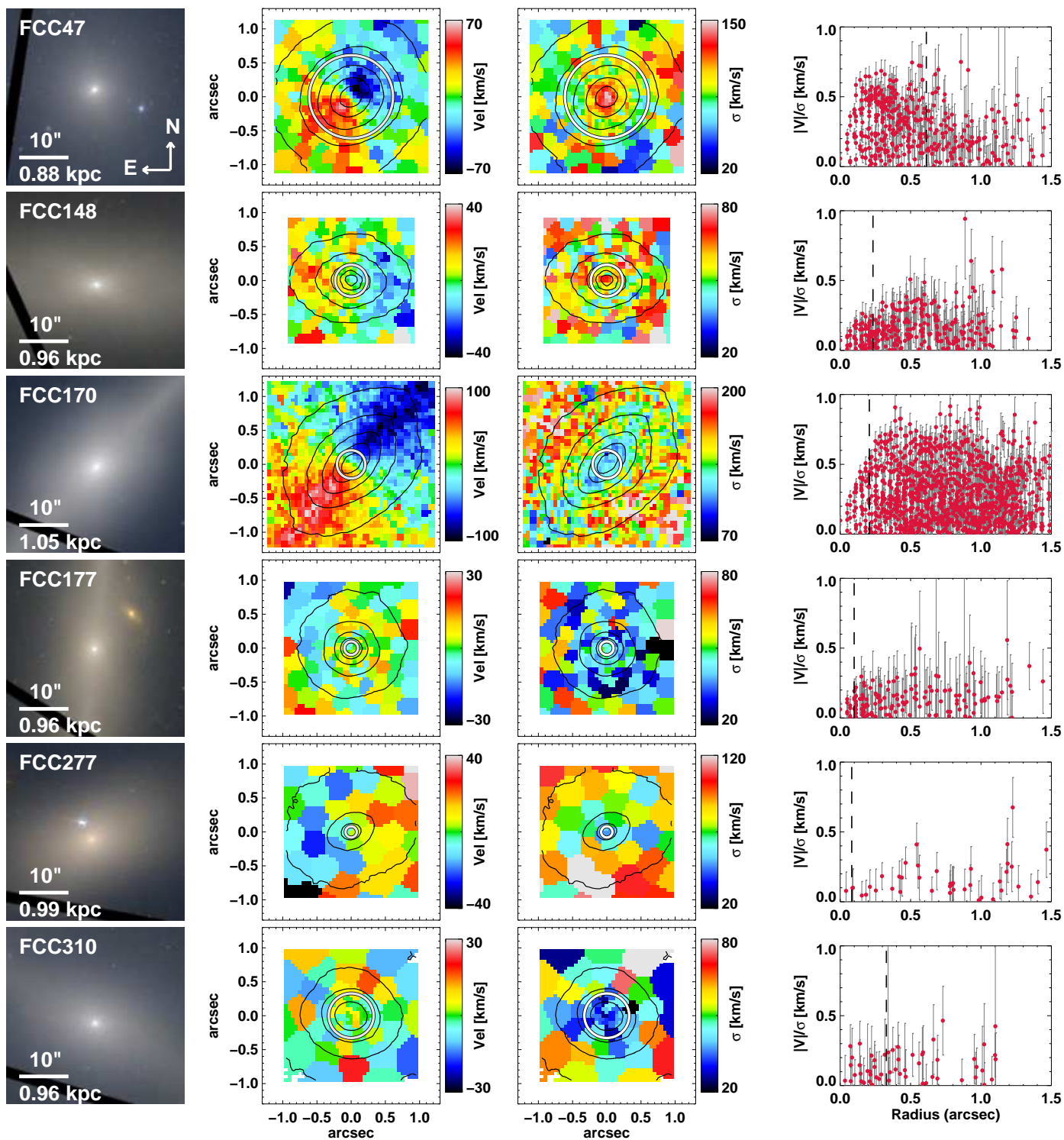}
\caption{\label{fig:stellar_kin_maps} {\it Left panels}: Colour-composite HST/ACS images of the 6 galaxies in our sample. {\it Middle panels}:  stellar mean velocity and velocity dispersion maps as derived from our SINFONI observations. Over-plotted are contours with constant surface brightness, as derived from our reconstructed  $K$-band images. The white circles illustrate the effective radius of the NSCs based on the inner Sersi\'{c} fits on HST $z$-band images \citep{turner12}. {\it Right panels}: $|V|/\sigma$ radial distribution of the bins. The vertical dashed lines illustrate the effective radii of the NSCs in $z$-band. All maps and images are oriented such that North is up and East is to the left.}
\end{figure*}

In Figure~\ref{fig:stellar_kin_maps} we present the mean stellar velocity and velocity dispersion maps of the nuclei of our sample of galaxies. We extracted these using pPXF \citep{ce04}, employing a library of seven template spectra of K and M giant stars, observed with the same instrumental setup. This set of stars, though limited in number due to the lack of a large and well calibrated near-IR stellar spectral library with sufficient spectral resolution, is shown to give fully consistent results in measuring the first and second stellar kinematics moments \citep{az08}. We fitted the wavelength range 2.1 -- 2.36~$\mu$m and masked the regions affected by strong near-IR sky lines \citep{rousselot00}. Additionally, we added a 6$^{th}$ order polynomial to the fit to compensate for template mismatch. We determined the best-fitting velocity and velocity dispersions, together with their uncertainties, as the bi-weighted mean and standard deviations of 100 Monte Carlo realisations. For consistency, we re-derived the kinematics maps of FCC\,277 binned to the same S/N$\sim$40 as the other galaxies.

Figure~\ref{fig:stellar_kin_maps} reveals a great variety of stellar kinematics in the nuclear regions of NSC hosting galaxies in the Fornax cluster. On each map we indicated with a white oval the extent of the effective radius of the inner Sersi\'{c} fit on HST $z$-band images of these galaxies \citep{turner12}. Our stellar kinematics maps are the sum of the NSC and the host galaxy kinematics. However, the NSCs light dominates the surface brightness profiles in the centres of nucleated galaxies \citep{cote06, turner12} . Therefore we can make a cautious assumption that the stellar kinematics that we observe inside the NSCs' effective radii is representative of the NSCs themselves (but see the discussion in Sect.~\ref{sec:lambda_sec}).

{\it FCC\,47} is the nearest galaxy in our sample and its NSC is most extended on the sky. The velocity map reveals a strong rotation (up to $\pm$~70 \kms) inside the NSC's $R_{eff}$ and a pronounced $\sigma$ peak reaching $\sim$~150 \kms\/ in the centre, indicative of the presence of a supermassive black hole.

{\it FCC\,148} shows a low level of rotation within the NSC's effective radius. The maximum velocity of $\pm$~35 \kms is reached outside of the NSC and overlaps with flattened isophotes, which is indicative of the presence of a nuclear disk. Indeed, \citet{turner12} showed that the best fit to this galaxy's light is obtained when a second large-scale disk component is added to the bulge and nucleus Sersi{\'c} models. The velocity dispersion map has a peak of $\sim$~70 \kms\/ in the centre, which is less prominent than the peak in the nucleus of FCC\,47.

{\it FCC\,170} hosts the most flattened NSC in our sample. The galaxy itself is viewed almost edge-on. The strong rotation in the central 2\arcsec $\times$ 2\arcsec reaches $\pm$~100 \kms. The $\sigma$ maps has a flattened drop in the centre.

{\it FCC\,177} does not show any ordered motions within the central 2\arcsec $\times$ 2\arcsec and the velocity dispersion map is flat.

The central kinematics of {\it FCC\,277} was studies in detail in \citet{az13}. The maximum rotation of $\pm$~25 \kms is reached outside of the NSC's effective radius and the velocity dispersion map has a distinct drop in the centre.

The stellar kinematics maps of {\it FCC\,310} is similar to FCC\,177: no clear observed rotation and a flat sigma map with a tentative central drop.

The right hand panels of Fig.~\ref{fig:stellar_kin_maps} show the radial distribution of the ordered vs. random motions ($|V|/\sigma$) inside our observed radii. The error bars reflect the excellent quality of our data, especially in the central areas occupied by the NSCs (indicated with vertical dashed lines). In all cases $|V|/\sigma < 0.6$ in the nuclear regions and are indicative of kinematically hot pressure supported systems. Traditionally, $|V|/\sigma$ has been used to examine the dynamical state of early-type galaxies \citep{davies83}. However, galaxies with very different mean velocity structures can have similar  $|V|/\sigma$ \citep{emsellem07}. Therefore, in order to explore the global velocity structure of galaxies using two-dimensional spatial kinematics, the specific stellar angular momentum, \lambdaRe\ \citep{emsellem07}, is largely used. Moreover, this quantity is closer to the true angular momentum of the systems.


\section{Specific angular momentum \lambdaRe\/}
\label{sec:lambda_sec}

In order to quantify the observed projected stellar angular momentum in our sample of 6 galactic nuclei we measured the specific stellar angular momentum \lambdaRe\/ following the standard equation \citep{emsellem07}:
\begin{equation}
\label{eq:sumLambda}
\lambda_R = \frac{\sum_{i=1}^{N_p} F_i R_i \left| V_i \right|}{\sum_{i=1}^{N_p} F_i R_i \sqrt{V_i^2+\sigma_i^2}} \, ,
\end{equation}
where $F_i$, $R_i$, $V_i$ and $\sigma_i$ are the flux, circular radius, velocity and velocity dispersion of the i$^{th}$ spatial bin and the sum runs on the $N_p$ bins.  We used the stellar kinematics maps presented in Section~\ref{sec:stellar_kin}. To define the extraction apertures we used the effective radius $R_{eff,z}$ of the inner Sersi\'{c} fit on HST/ACS $z$-band images \citep{turner12}, describing the NSCs in our sample of galaxies. The ellipticity (\epsE) and  position angle (PA) within that radius we estimated as the median \epsE \/ and PA of the isophotes in the range $0\farcs05 < r < R_{eff,z}$, based on the isophotal analysis of HST/ACS $z$-band images (Ferrarese et al., in preparation). 

 The ellipticities that we used might not be matching the true ellipticities of the NSCs. \citet{turner12} studied the 2-dimensional properties of galaxies and found that nuclei in brighter (and higher mass) galaxies are more flattened and may be more likely to contain edge-on disk-like components. However, they performed 2-dimensional surface brightness profile fitting only to galaxies with $B_{T} \geq 13.5,$ and therefore  FCC\,47, FCC\,170, and FCC\,177 were excluded. FCC\,148 and FCC\,310 required multi-component nucleus fits and thus no clean measurement was possible. The only galaxy from our sample that has a 2D fit is FCC\,277 and indeed the NSC measured ellipticity was 0.41, which is higher than the ellipticity measured over the  galaxy plus NSC inside the same radius (see Table~\ref{tab:lambda_all}). Taking into account that our observed stellar kinematics maps are the sum of these two components, and may be including nuclear disks additionally, we chose to define our extraction apertures based on the HST/ACS $z$-band total isophotal profiles. Even though  the stellar light in these nuclear regions is dominated by the NSCs, we take a cautious assumption that our \lambdaRe\/ values are representative of the galaxy nuclei, rather than of the NSCs only.

\begin{table}
\begin{center}
\caption{\label{tab:lambda_all} Basic properties of our galactic nuclei and their \lambdaRe\/. Columns (2) and (3) list the effective (half-light) radius $R_{eff,z}$ of the inner Sersi\'{c} model on HST/ACS $z$-band images from \citet{turner12} in arcseconds and parsecs, respectively. Columns (4) and (5) list the median isophotal ellipticity and position angle, respectively, inside the $R_{eff,z}$. Column (6) lists the measured \lambdaRe\/ of our nuclei.}
\begin{tabular}{c c c c c c}
\hline
\hline
Name  & $R_{eff,z}$ & $R_{eff,z}$ & \epsE & PA  & \lambdaRe\\
 & arcsec & pc & & degrees & \\
(1) & (2) & (3) & (4) & (5) & (6) \\
\hline
 FCC47  & 0.612  &  54  & 0.16 & 134  &  0.295  \\
FCC148 & 0.233  &  22  & 0.08 &   87  &  0.115  \\
FCC170 & 0.207  &  21  & 0.30 & 135  &  0.309  \\
FCC177 & 0.099  &   9  & 0.04 &   94  &  0.105  \\
FCC277 & 0.082  &   8  & 0.15 & 112  &  0.149  \\
FCC310 & 0.328  &  31  & 0.10 &   84  &  0.143  \\

\hline
\hline
\end{tabular}
\label{default}
\end{center}
\end{table}%

\begin{figure*}
{\includegraphics[width=\linewidth, trim=0.3cm 0cm 0.4cm 1.cm, clip]{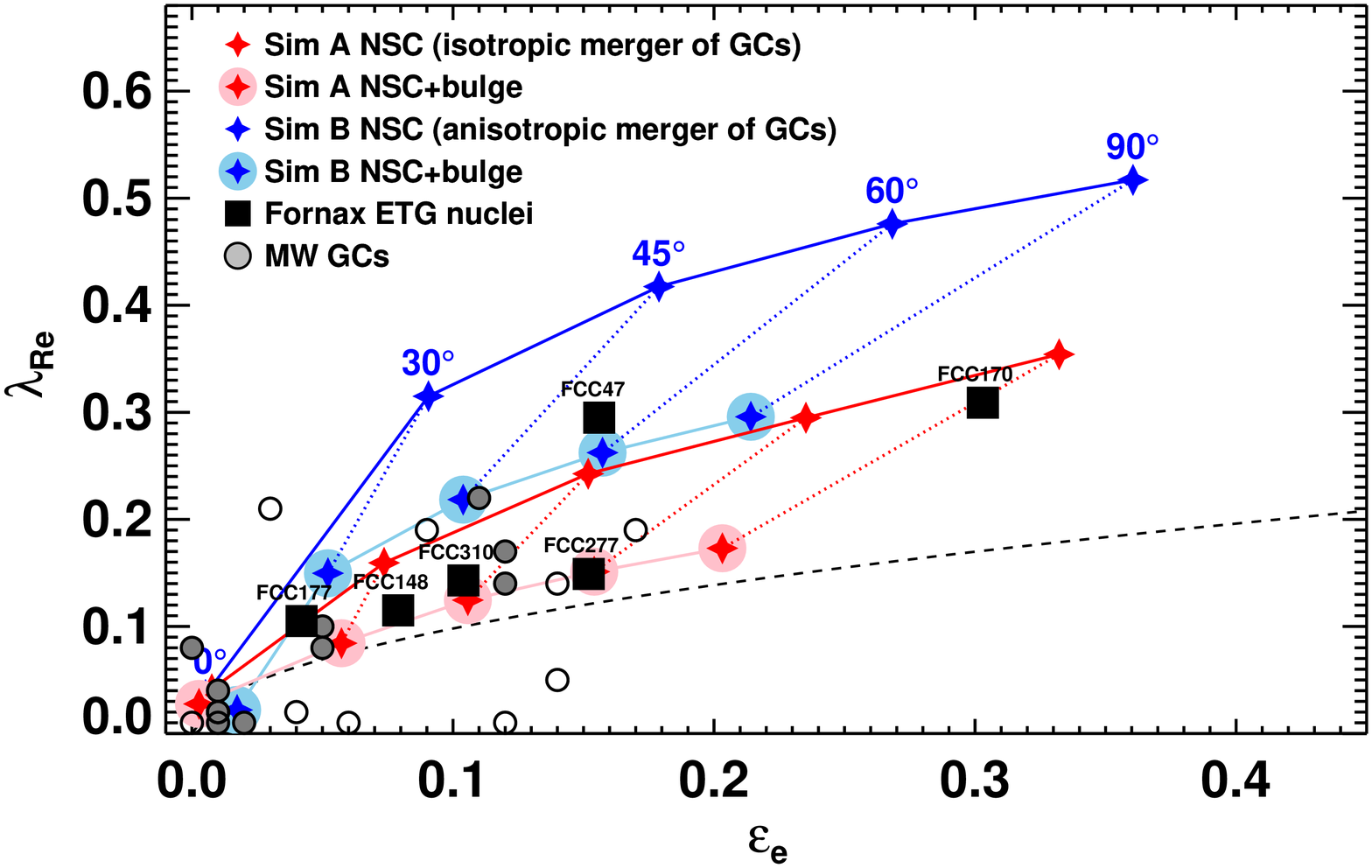}}
\caption{\label{fig:eps_lambda} The ellipticity vs. specific angular momentum \lambdaRe\/ of the NSCs in the 6 Fornax cluster galaxies (filled squares, uncertainties are smaller than the symbols) are compared with globular clusters in the Milky Way \citep[][open circles represent GCs where the kinematics data do not reach the half-light radius and \lambdaRe\/ is measured within the available maximum extent]{kamann18}.  The coloured symbols represent the values derived from the two simulated NSCs, described in Sect.~\ref{sec:sims}, the dotted line connect measurements with the same viewing angle. The dashed line denotes the separation line between slow- and fast-rotating galaxies \citep{emsellem11}.
}
\end{figure*}

In Figure~\ref{fig:eps_lambda}  we plotted our measurements of the specific stellar angular momentum \lambdaRe\/ inside the 6 nuclei in our sample vs. their mean ellipticities, \epsE, inside their NSCs' effective radii. As a comparison sample we used Milky Way globular clusters \citep{kamann18}, with ellipticities are taken from \citet[][2010 edition]{harris96}. This diagram is similar to the commonly used \lambdaRe--\epsE\/ diagram to classify galaxies into slow- and fast-rotators \citep{emsellem07, emsellem11, fb15_proc,cortese16,graham18}. 

\citet{kamann18} found that the amount of rotation of the studied globular clusters correlates with their ellipticities, in line with the findings of \citet{bellazzini12,bianchini13,kacharov14}. The six galaxy nuclei in our sample follow these trends as well and  all of them are above the division line between slow- and fast-rotators  \citep{emsellem11}. Thus they could be classified as mini fast-rotators based on this classification scheme. If we assumed that the results of the cosmological hydrodynamical simulations for the formation of early-type galaxies \citep[e.g.][]{naab14} were applicable to the formation of NSCs, then we would conclude that these 6 mini fast-rotator nuclei are the result of dissipative processes. This is consistent with the conclusions of e.g. \citet{turner12} and \citet{spengler17} that argue that nuclei in high-mass host galaxies (as the ones studied in our sample) grow through gas accretion triggered by wet mergers. 

Our  measurements of \lambdaRe\/ of the nuclei cannot be, however, directly compared to \lambdaRe\/ of globular clusters or galaxies due to the fact that the NSCs are embedded in the bulges of their host galaxies, as discussed above. Even though at first approximation the nuclear regions are dominated by the light of the NSCs, we cannot neglect the kinematical contribution of the over-imposed  bulge. Therefore, we used simulations of NSC formation to asses the contribution of the bulges to the observed nuclear \lambdaRe\/. 

\section{Comparison with simulated NSCs}
\label{sec:sims}

The simulated NSCs used here for comparison are similar to the NSC in the Milky Way (total mass of $\sim$ \ensuremath{10^{ 7 } $\Msun$ }) and are described in detail in \cite{antonini12,perets14, tsatsi17a}. They are formed in $N$-body simulations of the consecutive infall and merging of globular clusters (GCs) in a galactic centre, hosting a central super-massive black hole ($M_{\bullet}=$ \ensuremath{4 \times10^{ 6 } $\Msun$ }) and embedded in a spherical and initially non-rotating nuclear bulge of mass \ensuremath{10^{ 8 } $\Msun$ }. With the merging of globular clusters, the bulge gains angular momentum, though at a much lower level than the NSC \citep[see Fig. 1 of][]{tsatsi17a}.  The resulting effect on the stellar kinematics of the system NSC $+$ bulge is that it has a lower net rotation as compared to the NSC only.

For our comparison, we used NSCs formed in two different dry merger scenarios: Simulation A concerns a NSC formed by GCs that in-spiral to the galactic centre from randomly chosen orbital directions, and in Simulation B the NSC is formed by GCs that share a similar orbital direction of in-spiralling. We adopted a range of viewing angles: 0$^{\circ}$, 30$^{\circ}$, 45$^{\circ}$, 60$^{\circ}$, and 90$^{\circ}$ with respect to the total angular momentum vector, so that 90$^{\circ}$ corresponds to the projection where the rotation of the NSC is maximum in our line-of-sight (LOS). We then created ``SINFONI-like'' LOS kinematic maps in each projection, in a field of view covering one half-mass radius of each cluster ($\sim$10 pc) and a pixel size of 0.5 pc, from which we derived \lambdaRe\/ and \epsE\/ in a similar manner as in our Fornax nuclei described in Section~\ref{sec:lambda_sec}. 

Additionally, in order to understand how the galaxy host stars surrounding a NSC can contaminate its observed line-of-sight kinematics, we re-estimated \lambdaRe\/ and \epsE\/ for the simulated NSCs accounting for contamination from stars of their surrounding nuclear bulge in each projection. The kinematics of the simulated NSCs used here are described in more detail in \cite{tsatsi17a}. They found that a NSC that has formed through the merging of globular clusters, can have a significant rotation -- a fact that has so far been attributed mainly to gas infall -- and is compatible to the observed properties of the Milky Way NSC \citep{feldmeier14}.  In general, the measured \lambdaRe\/  of the bulge$+$NSC system will be a lower limit to the NSC \lambdaRe\/ only. The exact suppressive effect of a typical slowly rotating bugle \citep[e.g.][]{emsellem11,kin_dyn_0} depends on various parameters, i.e. structural properties, the misalignment of NSC compared to its bulge, and the viewing angle.

The derived \lambdaRe\/ and \epsE\/ for the two simulated NSCs are shown in Figure~\ref{fig:eps_lambda}. The first notable observation is that all measurements of the simulated NSCs can be classified as fast-rotating. In general \lambdaRe\/ increases with the viewing angle and when the measurement is decontaminated from the surrounding nuclear bulge. Thus, we can conclude that the measured \lambdaRe\/ inside the nuclei of the six Fornax galaxies  are lower limits to the specific angular momentum of their NSCs, under the assumption that their main contamination is by a slow-rotating component, such as a bulge. The observational fact that we see some of the NSCs embedded in rotating disks (i.e. FCC\,148, FCC\,170, and FCC\,277) has little effect on the measured \lambdaRe\/ inside the effective radii of the NSCs. This is because the light of the disks is extended, in contrast with the bulge and the NSC, which are both concentrated. As a result, the disk’s rotation contributes negligibly to the light-weighted measurement inside the NSC’s effective radii.

Second, the majority of the observed nuclei follow closely the predictions of Simulation A (red symbols), i.e. the case with isotropic GCs merging, and when we take into account the contamination by the nuclear bulge. The nucleus of FCC\,47, which is the most spatially extended and with the best resolved stellar kinematics available, overlaps with the contaminated Simulation B (blue symbols, GCs have fallen from a similar orbital direction). The detailed formation history of this NSC, including its stellar populations, are a subject of a forthcoming paper (Fahrion et al., {\it in preparation}). The \lambdaRe\/ and  \epsE\/ of the nucleus of FCC\,170 are inconsistent with the contaminated cases of any of the GC merger simulations. This is the most flattened nucleus in our sample and it is embedded in a metal rich nuclear stellar disk, which in turn is embedded in a bulge \citep{pinna19a},  indicating that a dissipative process might have played a significant role in its formation and growth \citep{guillard16}.

\section{Concluding remarks}
\label{sec:concl}

Nucleation is a wide spread phenomenon with galaxies of all types hosting nuclear star clusters (NSCs). However, we still have not  understood the mechanisms that govern their formation and growth. There are numerous studies that explore the photometric properties of NSCs. Spectroscopic data, however, are much more difficult to obtain, due to the small spatial extents that these stellar systems have and the small contrast with respect to their hosts, especially in the case of early-type galaxies.

In this paper we presented a new data set with integral-field spectroscopy of five nucleated galaxies in the Fornax cluster and included one more from our pilot study \citep{az13}.  We studied the stellar kinematics of the 6 nuclei and compared it with a sample of Milky Way globular clusters and a suit of $N$-body simulations that form NSCs via the in-spiralling of globular clusters.

We found that all studied nuclei exhibit varied stellar kinematics. Four out of six exhibit clear rotation, one shows a peak in velocity dispersion in its centres, indicative for super-massive black holes.  Other four nuclei display a velocity dispersion drop, including FCC\,277, for which we presented an orbit-based dynamical model in our pilot study and found that a significant rotation on counter rotating orbits is present.

When placed on the \lambdaRe\/--\epsE\/ diagram the 6 nuclei behave similarly to Milky Way GCs. Except of one case, the remaining five nuclei are consistent with the \lambdaRe\/--\epsE\/ of simulated NSCs (embedded in a contaminating nuclear bulge) that have formed via the merger of globular clusters. It has previously been suggested that the NSCs in higher mass galaxies, like the ones studied in this paper ($9.7 > \log(M_{dyn}/M_{\sun}) > 10.6$), form via dissipative processes with hints that some might also involve the merger of globular clusters. 

In this paper we showed that we cannot exclude the pure GC merging scenario as a viable path for the formation of NSCs. A caveat of our measurements is that they provide only lower limits to \lambdaRe\/ of the NSCs, due to the fact that we cannot directly separate the observed kinematical contribution of the NSC and the surrounding bulge. A way forward would be to obtain dynamical decompositions based on orbit-based Schwarzschild models and use these as priors to perform photometric decomposition of the IFU data, as recently discussed in \citet{zhu18b}.

\begin{acknowledgements}
We are grateful to the ESO astronomers who obtained the data presented in this paper in service mode operations at La Silla Paranal Observatory.
We acknowledge fruitful discussions with Alessandra Mastrobuono-Battisti, Glenn van de Ven, Harald Kuntschner, Monica Turner, Patrick C\^ot\'e, Eric Peng, Remco van den Bosch. We thank Laura Ferrarese for providing us with their results prior to publication. We also thank the referee for their valuable comments.

\end{acknowledgements}

\bibliographystyle{aa}
\bibliography{/Users/mlyubeno/Dropbox/sci/biblio/papers}

\begin{thebibliography}{54}
\expandafter\ifx\csname natexlab\endcsname\relax\def\natexlab#1{#1}\fi

\bibitem[{{Antonini} {et~al.}(2012){Antonini}, {Capuzzo-Dolcetta},
  {Mastrobuono-Battisti}, \& {Merritt}}]{antonini12}
{Antonini}, F., {Capuzzo-Dolcetta}, R., {Mastrobuono-Battisti}, A., \&
  {Merritt}, D. 2012, \apj, 750, 111

\bibitem[{{Barth} {et~al.}(2009){Barth}, {Strigari}, {Bentz}, {Greene}, \&
  {Ho}}]{barth09}
{Barth}, A.~J., {Strigari}, L.~E., {Bentz}, M.~C., {Greene}, J.~E., \& {Ho},
  L.~C. 2009, \apj, 690, 1031

\bibitem[{{Bellazzini} {et~al.}(2012){Bellazzini}, {Bragaglia}, {Carretta},
  {Gratton}, {Lucatello}, {Catanzaro}, \& {Leone}}]{bellazzini12}
{Bellazzini}, M., {Bragaglia}, A., {Carretta}, E., {et~al.} 2012, \aap, 538,
  A18

\bibitem[{{Bianchini} {et~al.}(2013){Bianchini}, {Varri}, {Bertin}, \&
  {Zocchi}}]{bianchini13}
{Bianchini}, P., {Varri}, A.~L., {Bertin}, G., \& {Zocchi}, A. 2013, \apj, 772,
  67

\bibitem[{{Blakeslee} {et~al.}(2009){Blakeslee}, {Jord{\'a}n}, {Mei},
  {C{\^o}t{\'e}}, {Ferrarese}, {Infante}, {Peng}, {Tonry}, \&
  {West}}]{blakeslee09}
{Blakeslee}, J.~P., {Jord{\'a}n}, A., {Mei}, S., {et~al.} 2009, \apj, 694, 556

\bibitem[{{B{\"o}ker} {et~al.}(2002){B{\"o}ker}, {Laine}, {van der Marel},
  {Sarzi}, {Rix}, {Ho}, \& {Shields}}]{boeker02}
{B{\"o}ker}, T., {Laine}, S., {van der Marel}, R.~P., {et~al.} 2002, \aj, 123,
  1389

\bibitem[{{Cappellari} {et~al.}(2006){Cappellari}, {Bacon}, {Bureau}, {Damen},
  {Davies}, {de Zeeuw}, {Emsellem}, {Falc{\'o}n-Barroso}, {Krajnovi{\'c}},
  {Kuntschner}, {McDermid}, {Peletier}, {Sarzi}, {van den Bosch}, \& {van de
  Ven}}]{cappellari06}
{Cappellari}, M., {Bacon}, R., {Bureau}, M., {et~al.} 2006, \mnras, 366, 1126

\bibitem[{{Cappellari} \& {Copin}(2003)}]{cc03}
{Cappellari}, M. \& {Copin}, Y. 2003, \mnras, 342, 345

\bibitem[{{Cappellari} \& {Emsellem}(2004)}]{ce04}
{Cappellari}, M. \& {Emsellem}, E. 2004, \pasp, 116, 138

\bibitem[{{Carson} {et~al.}(2015){Carson}, {Barth}, {Seth}, {den Brok},
  {Cappellari}, {Greene}, {Ho}, \& {Neumayer}}]{carson15}
{Carson}, D.~J., {Barth}, A.~J., {Seth}, A.~C., {et~al.} 2015, \aj, 149, 170

\bibitem[{{Cortese} {et~al.}(2016){Cortese}, {Fogarty}, {Bekki}, {van de
  Sande}, {Couch}, {Catinella}, {Colless}, {Obreschkow}, {Taranu}, {Tescari},
  {Barat}, {Bland-Hawthorn}, {Bloom}, {Bryant}, {Cluver}, {Croom},
  {Drinkwater}, {d'Eugenio}, {Konstantopoulos}, {Lopez-Sanchez}, {Mahajan},
  {Scott}, {Tonini}, {Wong}, {Allen}, {Brough}, {Goodwin}, {Green}, {Ho},
  {Kelvin}, {Lawrence}, {Lorente}, {Medling}, {Owers}, {Richards}, {Sharp}, \&
  {Sweet}}]{cortese16}
{Cortese}, L., {Fogarty}, L.~M.~R., {Bekki}, K., {et~al.} 2016, \mnras, 463,
  170

\bibitem[{{C{\^o}t{\'e}} {et~al.}(2006){C{\^o}t{\'e}}, {Piatek}, {Ferrarese},
  {Jord{\'a}n}, {Merritt}, {Peng}, {Ha{\c s}egan}, {Blakeslee}, {Mei}, {West},
  {Milosavljevi{\'c}}, \& {Tonry}}]{cote06}
{C{\^o}t{\'e}}, P., {Piatek}, S., {Ferrarese}, L., {et~al.} 2006, \apjs, 165,
  57

\bibitem[{{Davies} {et~al.}(1983){Davies}, {Efstathiou}, {Fall}, {Illingworth},
  \& {Schechter}}]{davies83}
{Davies}, R.~L., {Efstathiou}, G., {Fall}, S.~M., {Illingworth}, G., \&
  {Schechter}, P.~L. 1983, \apj, 266, 41

\bibitem[{{den Brok} {et~al.}(2014){den Brok}, {Peletier}, {Seth}, {Balcells},
  {Dominguez}, {Graham}, {Carter}, {Erwin}, {Ferguson}, {Goudfrooij},
  {Guzm{\'a}n}, {Hoyos}, {Jogee}, {Lucey}, {Phillipps}, \& {et
  al.}}]{denbrok14}
{den Brok}, M., {Peletier}, R.~F., {Seth}, A., {et~al.} 2014, \mnras, 445, 2385

\bibitem[{{Emsellem} {et~al.}(2011){Emsellem}, {Cappellari}, {Krajnovi{\'c}},
  {Alatalo}, {Blitz}, {Bois}, {Bournaud}, {Bureau}, {Davies}, {Davis}, {de
  Zeeuw}, {Khochfar}, {Kuntschner}, {Lablanche}, {McDermid}, {Morganti},
  {Naab}, {Oosterloo}, {Sarzi}, {Scott}, {Serra}, {van de Ven}, {Weijmans}, \&
  {Young}}]{emsellem11}
{Emsellem}, E., {Cappellari}, M., {Krajnovi{\'c}}, D., {et~al.} 2011, \mnras,
  414, 888

\bibitem[{{Emsellem} {et~al.}(2007){Emsellem}, {Cappellari}, {Krajnovi{\'c}},
  {van de Ven}, {Bacon}, {Bureau}, {Davies}, {de Zeeuw}, {Falc{\'o}n-Barroso},
  {Kuntschner}, {McDermid}, {Peletier}, \& {Sarzi}}]{emsellem07}
{Emsellem}, E., {Cappellari}, M., {Krajnovi{\'c}}, D., {et~al.} 2007, \mnras,
  379, 401

\bibitem[{{Falc{\'o}n-Barroso} {et~al.}(2015){Falc{\'o}n-Barroso}, {Lyubenova},
  \& {van de Ven}}]{fb15_proc}
{Falc{\'o}n-Barroso}, J., {Lyubenova}, M., \& {van de Ven}, G. 2015, in IAU
  Symposium, Vol. 311, Galaxy Masses as Constraints of Formation Models, ed.
  M.~{Cappellari} \& S.~{Courteau}, 78--81

\bibitem[{{Falc{\'o}n-Barroso} {et~al.}(2017){Falc{\'o}n-Barroso}, {Lyubenova},
  {van de Ven}, {Mendez-Abreu}, {Aguerri}, {Garc{\'{\i}}a-Lorenzo},
  {Bekerait{\'e}}, {S{\'a}nchez}, {Husemann}, {Garc{\'{\i}}a-Benito}, {Mast},
  {Walcher}, {Zibetti}, {Barrera-Ballesteros}, {Galbany},
  {S{\'a}nchez-Bl{\'a}zquez}, {Singh}, {van den Bosch}, {Wild}, {Zhu},
  {Bland-Hawthorn}, {Cid Fernandes}, {de Lorenzo-C{\'a}ceres}, {Gallazzi},
  {Gonz{\'a}lez Delgado}, {Marino}, {M{\'a}rquez}, {P{\'e}rez}, {P{\'e}rez},
  {Roth}, {Rosales-Ortega}, {Ruiz-Lara}, {Wisotzki}, {Ziegler}, \& {Califa
  Collaboration}}]{kin_dyn_0}
{Falc{\'o}n-Barroso}, J., {Lyubenova}, M., {van de Ven}, G., {et~al.} 2017,
  \aap, 597, A48

\bibitem[{{Feldmeier} {et~al.}(2014){Feldmeier}, {Neumayer}, {Seth},
  {Sch{\"o}del}, {L{\"u}tzgendorf}, {de Zeeuw}, {Kissler-Patig}, {Nishiyama},
  \& {Walcher}}]{feldmeier14}
{Feldmeier}, A., {Neumayer}, N., {Seth}, A., {et~al.} 2014, \aap, 570, A2

\bibitem[{{Ferguson}(1989)}]{ferguson89}
{Ferguson}, H.~C. 1989, \aj, 98, 367

\bibitem[{{Ferrarese} {et~al.}(2006){Ferrarese}, {C{\^o}t{\'e}}, {Dalla
  Bont{\`a}}, {Peng}, {Merritt}, {Jord{\'a}n}, {Blakeslee}, {Ha{\c s}egan},
  {Mei}, {Piatek}, {Tonry}, \& {West}}]{ferrarese06}
{Ferrarese}, L., {C{\^o}t{\'e}}, P., {Dalla Bont{\`a}}, E., {et~al.} 2006,
  \apjl, 644, L21

\bibitem[{{Georgiev} \& {B{\"o}ker}(2014)}]{georgiev14}
{Georgiev}, I.~Y. \& {B{\"o}ker}, T. 2014, \mnras, 441, 3570

\bibitem[{{Georgiev} {et~al.}(2016){Georgiev}, {B{\"o}ker}, {Leigh},
  {L{\"u}tzgendorf}, \& {Neumayer}}]{georgiev16}
{Georgiev}, I.~Y., {B{\"o}ker}, T., {Leigh}, N., {L{\"u}tzgendorf}, N., \&
  {Neumayer}, N. 2016, \mnras, 457, 2122

\bibitem[{{Graham} \& {Spitler}(2009)}]{graham09}
{Graham}, A.~W. \& {Spitler}, L.~R. 2009, \mnras, 397, 2148

\bibitem[{{Graham} {et~al.}(2018){Graham}, {Cappellari}, {Li}, {Mao},
  {Bershady}, {Bizyaev}, {Brinkmann}, {Brownstein}, {Bundy}, {Drory}, {Law},
  {Pan}, {Thomas}, {Wake}, {Weijmans}, {Westfall}, \& {Yan}}]{graham18}
{Graham}, M.~T., {Cappellari}, M., {Li}, H., {et~al.} 2018, \mnras, 477, 4711

\bibitem[{{Guillard} {et~al.}(2016){Guillard}, {Emsellem}, \&
  {Renaud}}]{guillard16}
{Guillard}, N., {Emsellem}, E., \& {Renaud}, F. 2016, \mnras, 461, 3620

\bibitem[{{Harris}(1996)}]{harris96}
{Harris}, W.~E. 1996, \aj, 112, 1487

\bibitem[{{Hartmann} {et~al.}(2011){Hartmann}, {Debattista}, {Seth},
  {Cappellari}, \& {Quinn}}]{hartmann11}
{Hartmann}, M., {Debattista}, V.~P., {Seth}, A., {Cappellari}, M., \& {Quinn},
  T.~R. 2011, \mnras, 418, 2697

\bibitem[{{Jahnke} \& {Macci{\`o}}(2011)}]{jahnke11}
{Jahnke}, K. \& {Macci{\`o}}, A.~V. 2011, \apj, 734, 92

\bibitem[{{Kacharov} {et~al.}(2014){Kacharov}, {Bianchini}, {Koch}, {Frank},
  {Martin}, {van de Ven}, {Puzia}, {McDonald}, {Johnson}, \&
  {Zijlstra}}]{kacharov14}
{Kacharov}, N., {Bianchini}, P., {Koch}, A., {et~al.} 2014, \aap, 567, A69

\bibitem[{{Kamann} {et~al.}(2018){Kamann}, {Husser}, {Dreizler}, {Emsellem},
  {Weilbacher}, {Martens}, {Bacon}, {den Brok}, {Giesers}, {Krajnovi{\'c}},
  {Roth}, {Wendt}, \& {Wisotzki}}]{kamann18}
{Kamann}, S., {Husser}, T.-O., {Dreizler}, S., {et~al.} 2018, \mnras, 473, 5591

\bibitem[{{Krajnovi{\'c}} {et~al.}(2011){Krajnovi{\'c}}, {Emsellem},
  {Cappellari}, {Alatalo}, {Blitz}, {Bois}, {Bournaud}, {Bureau}, {Davies},
  {Davis}, {de Zeeuw}, {Khochfar}, {Kuntschner}, {Lablanche}, \& {et
  al.}}]{krajnovic11}
{Krajnovi{\'c}}, D., {Emsellem}, E., {Cappellari}, M., {et~al.} 2011, \mnras,
  414, 2923

\bibitem[{{Krist}(1995)}]{krist95}
{Krist}, J. 1995, in Astronomical Society of the Pacific Conference Series,
  Vol.~77, Astronomical Data Analysis Software and Systems IV, ed. R.~A.
  {Shaw}, H.~E. {Payne}, \& J.~J.~E. {Hayes}, 349

\bibitem[{{Kuntschner}(2000)}]{kun00}
{Kuntschner}, H. 2000, \mnras, 315, 184

\bibitem[{{Lyubenova} {et~al.}(2008){Lyubenova}, {Kuntschner}, \&
  {Silva}}]{az08}
{Lyubenova}, M., {Kuntschner}, H., \& {Silva}, D.~R. 2008, \aap, 485, 425

\bibitem[{{Lyubenova} {et~al.}(2013){Lyubenova}, {van den Bosch},
  {C{\^o}t{\'e}}, {Kuntschner}, {van de Ven}, {Ferrarese}, {Jord{\'a}n},
  {Infante}, \& {Peng}}]{az13}
{Lyubenova}, M., {van den Bosch}, R.~C.~E., {C{\^o}t{\'e}}, P., {et~al.} 2013,
  \mnras, 431, 3364

\bibitem[{{Mihos} \& {Hernquist}(1994)}]{mihos94}
{Mihos}, J.~C. \& {Hernquist}, L. 1994, \apjl, 437, L47

\bibitem[{{Naab} {et~al.}(2014){Naab}, {Oser}, {Emsellem}, {Cappellari},
  {Krajnovi{\'c}}, {McDermid}, {Alatalo}, {Bayet}, {Blitz}, {Bois}, {Bournaud},
  {Bureau}, {Crocker}, {Davies}, {Davis}, {de Zeeuw}, {Duc}, {Hirschmann},
  {Johansson}, {Khochfar}, {Kuntschner}, {Morganti}, {Oosterloo}, {Sarzi},
  {Scott}, {Serra}, {Ven}, {Weijmans}, \& {Young}}]{naab14}
{Naab}, T., {Oser}, L., {Emsellem}, E., {et~al.} 2014, \mnras, 444, 3357

\bibitem[{{Neumayer} {et~al.}(2011){Neumayer}, {Walcher}, {Andersen},
  {S{\'a}nchez}, {B{\"o}ker}, \& {Rix}}]{neumayer11}
{Neumayer}, N., {Walcher}, C.~J., {Andersen}, D., {et~al.} 2011, \mnras, 413,
  1875

\bibitem[{{Perets} \& {Mastrobuono-Battisti}(2014)}]{perets14}
{Perets}, H.~B. \& {Mastrobuono-Battisti}, A. 2014, \apjl, 784, L44

\bibitem[{{Pinna} {et~al.}(2019){Pinna}, {Falc{\'o}n-Barroso}, {Martig},
  {Sarzi}, {Coccato}, {Iodice}, {Corsini}, {de Zeeuw}, {Gadotti}, {Leaman},
  {Lyubenova}, {McDermid}, {Minchev}, {Morelli}, {van de Ven}, \&
  {Viaene}}]{pinna19a}
{Pinna}, F., {Falc{\'o}n-Barroso}, J., {Martig}, M., {et~al.} 2019, \aap, 623,
  A19

\bibitem[{{Rossa} {et~al.}(2006){Rossa}, {van der Marel}, {B{\"o}ker},
  {Gerssen}, {Ho}, {Rix}, {Shields}, \& {Walcher}}]{rossa06}
{Rossa}, J., {van der Marel}, R.~P., {B{\"o}ker}, T., {et~al.} 2006, \aj, 132,
  1074

\bibitem[{{Rousselot} {et~al.}(2000){Rousselot}, {Lidman}, {Cuby}, {Moreels},
  \& {Monnet}}]{rousselot00}
{Rousselot}, P., {Lidman}, C., {Cuby}, J.-G., {Moreels}, G., \& {Monnet}, G.
  2000, \aap, 354, 1134

\bibitem[{{Seth} {et~al.}(2008){Seth}, {Blum}, {Bastian}, {Caldwell}, \&
  {Debattista}}]{seth08}
{Seth}, A.~C., {Blum}, R.~D., {Bastian}, N., {Caldwell}, N., \& {Debattista},
  V.~P. 2008, \apj, 687, 997

\bibitem[{{Seth} {et~al.}(2006){Seth}, {Dalcanton}, {Hodge}, \&
  {Debattista}}]{seth06}
{Seth}, A.~C., {Dalcanton}, J.~J., {Hodge}, P.~W., \& {Debattista}, V.~P. 2006,
  \aj, 132, 2539

\bibitem[{{Silk} \& {Rees}(1998)}]{silk98}
{Silk}, J. \& {Rees}, M.~J. 1998, \aap, 331, L1

\bibitem[{{Spengler} {et~al.}(2017){Spengler}, {C{\^o}t{\'e}}, {Roediger},
  {Ferrarese}, {S{\'a}nchez-Janssen}, {Toloba}, {Liu}, {Guhathakurta},
  {Cuillandre}, {Gwyn}, {Zirm}, {Mu{\~n}oz}, {Puzia}, {Lan{\c c}on}, {Peng},
  {Mei}, \& {Powalka}}]{spengler17}
{Spengler}, C., {C{\^o}t{\'e}}, P., {Roediger}, J., {et~al.} 2017, \apj, 849,
  55

\bibitem[{{Tremaine} {et~al.}(1975){Tremaine}, {Ostriker}, \&
  {Spitzer}}]{tremaine75}
{Tremaine}, S.~D., {Ostriker}, J.~P., \& {Spitzer}, Jr., L. 1975, \apj, 196,
  407

\bibitem[{{Tsatsi} {et~al.}(2017){Tsatsi}, {Mastrobuono-Battisti}, {van de
  Ven}, {Perets}, {Bianchini}, \& {Neumayer}}]{tsatsi17a}
{Tsatsi}, A., {Mastrobuono-Battisti}, A., {van de Ven}, G., {et~al.} 2017,
  \mnras, 464, 3720

\bibitem[{{Turner} {et~al.}(2012){Turner}, {C{\^o}t{\'e}}, {Ferrarese},
  {Jord{\'a}n}, {Blakeslee}, {Mei}, {Peng}, \& {West}}]{turner12}
{Turner}, M.~L., {C{\^o}t{\'e}}, P., {Ferrarese}, L., {et~al.} 2012, \apjs,
  203, 5

\bibitem[{{Walcher} {et~al.}(2005){Walcher}, {van der Marel}, {McLaughlin},
  {Rix}, {B{\"o}ker}, {H{\"a}ring}, {Ho}, {Sarzi}, \& {Shields}}]{walcher05}
{Walcher}, C.~J., {van der Marel}, R.~P., {McLaughlin}, D., {et~al.} 2005,
  \apj, 618, 237

\bibitem[{{Wegner} {et~al.}(2003){Wegner}, {Bernardi}, {Willmer}, {da Costa},
  {Alonso}, {Pellegrini}, {Maia}, {Chaves}, \& {Rit{\'e}}}]{wegner03}
{Wegner}, G., {Bernardi}, M., {Willmer}, C.~N.~A., {et~al.} 2003, \aj, 126,
  2268

\bibitem[{{Weijmans} {et~al.}(2014){Weijmans}, {de Zeeuw}, {Emsellem},
  {Krajnovi{\'c}}, {Lablanche}, {Alatalo}, {Blitz}, {Bois}, {Bournaud},
  {Bureau}, {Cappellari}, {Crocker}, {Davies}, {Davis}, {Duc}, {Khochfar},
  {Kuntschner}, {McDermid}, {Morganti}, {Naab}, {Oosterloo}, {Sarzi}, {Scott},
  {Serra}, {Verdoes Kleijn}, \& {Young}}]{weijmans14}
{Weijmans}, A.-M., {de Zeeuw}, P.~T., {Emsellem}, E., {et~al.} 2014, \mnras,
  444, 3340

\bibitem[{{Zhu} {et~al.}(2018){Zhu}, {van de Ven}, {M{\'e}ndez-Abreu}, \&
  {Obreja}}]{zhu18b}
{Zhu}, L., {van de Ven}, G., {M{\'e}ndez-Abreu}, J., \& {Obreja}, A. 2018,
  \mnras, 479, 945

\end{thebibliography}

\end{document}